# A Quantitative Investigation of $CO_2$ Sequestration by Mineral Carbonation

Muneer Mohammad, Student Member, *IEEE* Mehrdad Ehsani, Fellow, *IEEE*

*Abstract*—Anthropogenic activities have led to a substantial increase in carbon dioxide ($CO_2$), a greenhouse gas (GHG), contributing to heightened concerns of global warming. In the last decade alone $CO_2$ emissions increased by 2.0 ppm/yr. globally. In the year 2009, United States and China contributed up to 43.4% of global $CO_2$ emissions. $CO_2$ capture and sequestration have been recognized as promising solutions to mitigate $CO_2$ emissions from fossil fuel based power plants. Typical techniques for carbon capture include post-combustion capture, pre-combustion capture and oxy-combustion capture, which are under active research globally. Mineral carbonation has been investigated as a suitable technique for long term storage of $CO_2$. Sequestration is a highly energy intensive process and the additional energy is typically supplied by the power plant itself. This leads to a reduction in net amount of $CO_2$ captured because of extra $CO_2$ emitted. This paper presents a quantitative analysis of the energy consumption during sequestration process for a typical 1GW pulverized coal and a 1GW natural gas based power plant. Furthermore, it has been established that the present day sequestration methods and procedures are not viable to achieve the goal of carbon sequestration.

*Index Terms*—Mineral Carbonation, $CO_2$ Sequestration, Calcium Silicates

## I. INTRODUCTION

AMONG the shocking warnings of severe weather alarms and globally rising temperatures, politicians, scientists, engineers, and others are searching for ways to reduce the rising threat of climate change. Currently, fossil fuels are the main source satisfying the global primary energy demand, and will remain or the rest of 21$^{st}$ century [1]. Approximately 85% of the world's energy needs are supplied by fossil fuels because of their low cost, existing reliable technology for energy production, and availability. As it is shown in Fig. 1, approximately 70% of United States' electricity needs are met by coal and natural gas power plants [2]. Presently, renewable energy power plants based on solar and wind are too sporadic to satisfy major electricity needs and transmission and infrastructure issues hinder expansion of their share in the market. Nuclear plants deliver base-load electricity, but the future of nuclear energy is plagued by several techno-economic and social issues. This leaves natural gas and coal as the two resources that could play substantial roles in the near-term addition of electrical generating capacity. The ignition of fossil fuels produces carbon dioxide ($CO_2$), which is the major greenhouse gas (GHG); many climatologists believe that emission of this GHG into the atmosphere is the major cause of the climate change.

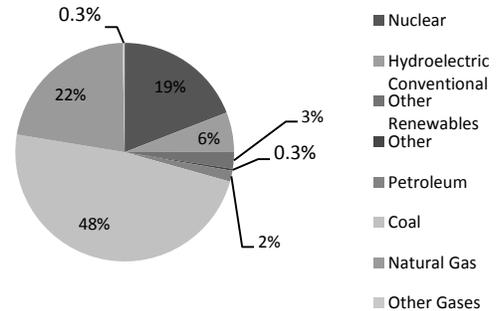

Fig. 1. 2007 U.S. Electric Power Industry Net Generation

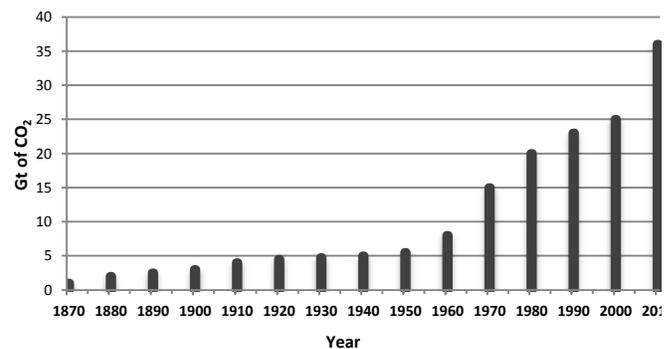

Fig. 2. Net $CO_2$ emission from burning fossil fuels

Fig. 2 depicts the global $CO_2$ emissions from burning fossil fuels during the last century. In the last 50 years alone the $CO_2$ emissions have risen by 300%. One of the methods for mitigating potential global climate change due to anthropogenic emissions of $CO_2$ into atmosphere is to capture $CO_2$ from burning fossil fuels, and to store it in oceanic or geologic reservoirs [1, 3, 4]. This is not the only solution, but the development of carbon capture and sequestration technology, which has accelerated significantly in the last decade, might play a vital role in addressing this issue. Carbon sequestration is a critical technology for decreasing greenhouse gas GHG) emissions from fossil fuel power plants. Capture of approximately 90% of emissions has significant progressive impacts on the technology, plant performance, and project economics. However, it also presents challenges for the first movers who implement the technology [1, 5].

All $CO_2$ sequestration methods consist of two steps. In the first step, the carbon dioxide is captured and separated from the flue gas or the fuel. The $CO_2$, having been isolated, is then stored in a reservoir [1, 5-7]. Active research is being undertaken in both these areas. Some of these technologies are at an early stage of development. Separation of $CO_2$ from the captured flue gas is required to avoid storage of huge volumes of $N_2$. The separation is typically established using absorption with monoethanolamine followed by stripping with steam. Absorption technology is energy-intensive and usually accounts for two-thirds to threequarters of the total sequestration costs.

$CO_2$ capture and isolation from power plants can be achieved via three prevalent technologies:

1. Flue gas separation: This method is based on chemical absorption. In a power plant employing this technology, the flue gas is passed through a solvent in a packed absorber column, wherein $CO_2$ is absorbed by the solvent on account of some chemical bond formation. The solvent then passes into a regenerator unit, where $CO_2$ is extracted from it and may be put into commercial uses or stored in reservoirs. Although solvent based extraction of $CO_2$ involves additional cost, this technology can be fitted onto the existing power plants easily. This technique is a prime example of post combustion $CO_2$ capture [4].

2. Pre-combustion carbon capture: This method involves capturing $CO_2$ before fuel is burnt in the plants. As the captured $CO_2$ has a higher concentration than the previous method, these separation methods turn out to be more efficient. Physical solvents which extract $CO_2$ via absorption are more suited since $CO_2$ is obtained at high pressures when syngas exits shifting converters. Other techniques such as polymer-based membranes and chemical looping combustion are in research stages. Pre-combustion carbon capture is more suited for integrated coal gasification combined cycle power plants (IGCC) wherein coal undergoes gasification to form syngas (CO + Hydrogen). Hydrogen is utilized as the main fuel while CO can be further converted to $CO_2$ easily to complete carbon capture. Cost is a major drawback as IGCC plants produce power at a higher cost than conventional pulverized coal powered plants. Also, this process could be used in industrial settings which involve production of syngas from coal.

3. Oxycombustion: This involves combustion of fossil fuel in an atmosphere enriched with oxygen. Again, the intent is to obtain a higher concentration of $CO_2$ in the byproduct of combustion. The higher concentration of $CO_2$ in the flue gas mixture makes separation process less energy intensive. In this method, pure oxygen is usually obtained via an air separation unit which consumes about 15% of a power plant's generated power. It may also prove to be a cost hindrance since cryogenic systems are required for oxygen separation.

Once $CO_2$ has been captured it needs to be stored. Various techniques exist for $CO_2$ storage. Mineral Carbonation seems to be a promising technique for $CO_2$ storage [1, 5, 7]. This technique involves reacting $CO_2$ with reactive calcium or magnesium ores (silicates/oxides) and obtaining calcium/magnesium carbonates, thereby storing $CO_2$ in these compounds. The carbonates could be further used in the industry or stored as landfills. The main motivation for the mineralization technique is that since carbonates have lower energy levels than $CO_2$, carbon dioxide to carbonates conversion is an exothermic reaction, which does not need energy inputs theoretically. However, for better kinetics, energy is provided. This storage methodology is a permanent solution against geological/ocean bed sequestration, which could lead to $CO_2$ leakage [3,7, 8].

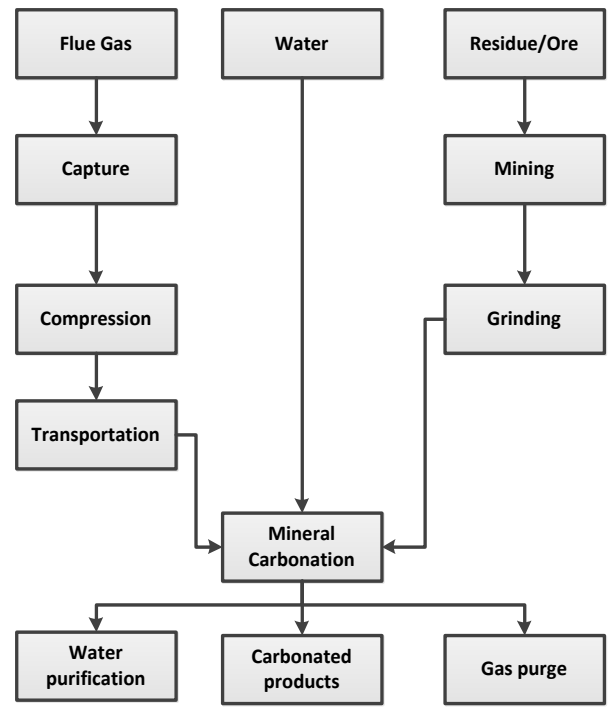

Fig. 3. Procedure of mineral carbonation process

Possible feedstock for mineral $CO_2$ sequestration include primary Ca/Mg-silicates, such as wollastonite ($CaSiO_3$) [9] and olivine ($Mg_2SiO_4$) and industrial residues, such as steel slag [10] and waste cement [11]. The proposed procedure of the entire $CO_2$ sequestration based on the mineral carbonation includes capture of flue gas, compression, transportation, mining, and grinding is presented in Fig. 2. The carbonation reaction is usually very slow; enhancement of this process in mineral carbonation is possible by grinding the feedstock, elevating the temperature and $CO_2$ pressure. However this will consume energy and decrease the net amount of $CO_2$ sequestered because of the additional $CO_2$ emissions caused. But the mineral carbonation reaction generates usable heat which can be utilized for producing part of the energy requirements. Therefore energy and efficiency analysis play an important role which is the main goal of this paper.

As mentioned earlier, the major cost of the carbon sequestration is the energy requirement; therefore in this paper energy analysis will be covered. Because the purpose of carbon capture sequestration is to decrease the amount of $CO_2$ emitted into the air, the energy used in the process itself shouldn't cause an extra $CO_2$ emission. Therefore a trade-off between the sequestration and the energy requirement for this process should be done. So we need to minimize the emitted $CO_2$ by preventing increase in generation capacity of the fossil fuel power plant for the purpose of sequestration. This paper discusses a scenario disturbing picture wherein the amount of $CO_2$ produced during sequestration is comparable to the $CO_2$ generated from the power plant. Furthermore, the energy consumed during carbon capture will be discussed in detail. This energy either needs to be provided externally or can be apportioned from the existing power plant.

## II. Literature Review

Carbon capture and carbon storage techniques have been under active research for about two decades now. Reducing $CO_2$ emissions by sequestering $CO_2$ from fossil fuel based power plants and storing it in the oceans was proposed by Marchetti in as early as 1977 [12]. Academic interest in this field took flight in the 1980s and 1990s [13-19] leading to International Conference on Carbon Dioxide Removal (ICCDR-1) in 1992 where researchers from all around the world met and discussed ideas on $CO_2$ capture and storage. The conference opened up active research in this field tremendously. Around 1991, International Energy Agency (IEA) established a mandate to research and develop techniques for capture and storage of $CO_2$ and help enhance international awareness on climate change [20]. Japan has been at the forefront in this field and has established an international organization for $CO_2$ fixation and utilization RITE (Research Institute of Innovative Technology for the Earth), which actively focuses on developing innovative environmental technologies for $CO_2$ sequestration.

Popular techniques for capturing $CO_2$ from power plants include post combustion $CO_2$ capture from flue gas, which involves isolating $CO_2$ after the fuel is burnt. Pre-combustion isolation of $CO_2$ is the process of converting fuel to syngas and extracting $CO_2$ via the water gas shift process. Oxycombustion involves fuel combustion in a highly oxygen enriched environment and $CO_2$ is captured from flue gases after fuel combustion [4, 5].

Once $CO_2$ has been captured and isolated, it is stored in carbon sinks. Ocean sequestration technique involves speeding up the natural process of $CO_2$ absorption in the oceans [21-24]. A large amount of $CO_2$ can be theoretically sequestered in the oceans. However this technology has a lot of environmental concerns [25] primarily with increasing acidification of the oceans. Another viable option for storage is storing $CO_2$ in geological formations such as deep saline formations, coal beds, and depleted oil and gas reservoirs. Geological storage is being researched as it presents a cost effective storage alternative for large amounts of $CO_2$ storage. Especially, deep saline aquifers have very high potential storage capacities for $CO_2$ because they are wide spread and available [22]. In contrast to oil and gas reservoirs, they do not require special structural trap geometries [26]. However, uncertainties in long term storage integrity and cost of the storage techniques still pose a stumbling block to this process. Human activities such as drilling, geological fractures, and structural leaks all contribute to the increased risk of $CO_2$ leakage.

Carbon dioxide storage by mineral carbonation is an alternative to storage in oceans or reservoirs [27]. It offers some significant advantages over $CO_2$ storage in geological reservoirs and oceans. Mineral carbonation provides a permanent $CO_2$ storage solution by conversion of $CO_2$ [28] into carbonates since chemical sequestration reaction produces thermodynamically stable products. It eliminates potential hazards of $CO_2$ leakage encountered with storage in geological formations or acidification of oceans [29-31]. An inherent benefit of carbonation process is that the carbonation reaction is an exothermic reaction, which theoretically implies no energy input for the chemical reaction [32]. However, naturally occurring reactions are slow, and different techniques to speed up the kinetics in efficient ways have been researched into. Carbonation processes with olivine and serpentine ores have been studied in detail [33]. Mineral carbonation requires abundant quantities of Ca or Mg silicate ores. The advantage that this process offers is the wide spread availability of minerals [27] such as Wollastonite ($CaSiO_3$) or Olivine ($Mg_2SiO_4$).

Huijen et.al [9] provide a method to carry out energy efficient $CO_2$ carbonation by aqueous mineralization of wollastonite ore [9]. The energy analysis has been carried further in this paper. The energy utilized for a 1000 MW pulverized coal based power plant adopting $CO_2$ sequestration will be determined. A similar analysis has been carried out for a 1000 MW natural gas based power plant incorporating $CO_2$ sequestration to isolate and sink $CO_2$.

## III. Case Study

A 1000 MW pulverized coal (PC) power plant and a 1000 MW natural gas powered plant have been considered to compute the $CO_2$ emissions.

There are three current technologies for pulverized coal plant. The ultra-supercritical pulverized coal plant has been known to have the highest efficiency. The efficiencies of these power plants are commonly estimated to be around 44% and typical natural gas power plants have efficiencies around 55%.

The calorific value of Bituminous coal is known to be in the range of 17 MJ/kg to 23.5 MJ/kg while the higher-grade anthracite coal has a higher calorific value of 32 MJ/kg to 35 MJ/kg. Bituminous coal (calorific value 20 MJ/kg) has been used as fuel for PC, given the relative abundance of bituminous coal as compared to anthracite. Also, calorific value of methane (natural gas) has been selected to be 55 MJ/kg. With the above specifications for pulverized coal plant and natural gas based power plant, the daily requirement of fuel (bituminous coal/natural gas) has been computed for the respective plants.

These numbers are later used to perform a molar analysis and the amount of daily carbon dioxide emissions from both the plants has been computed.

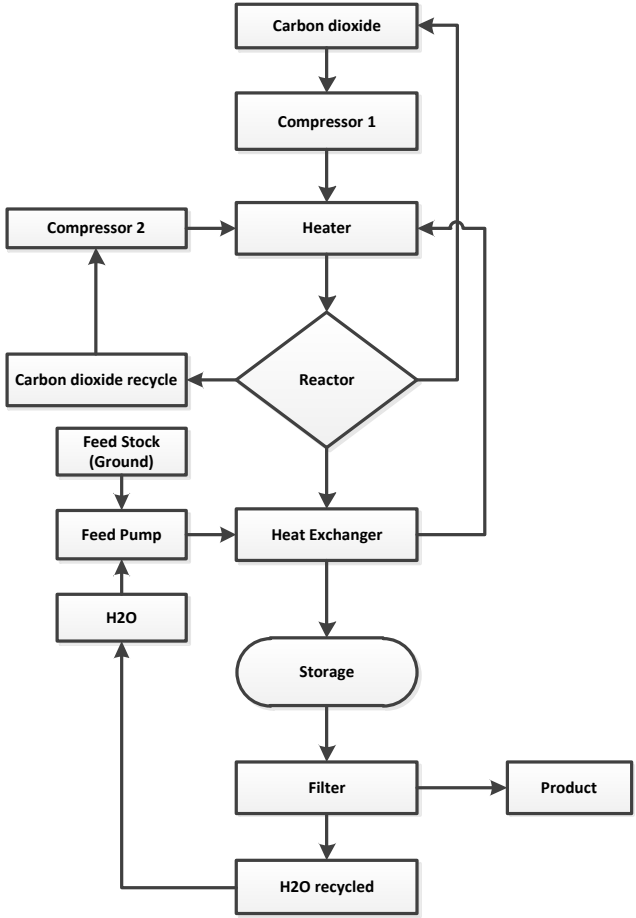

Fig. 4. Mineral carbonation process flow

In the next step, energy efficiency of mineral carbonation process has been calculated. The calcium mineral (wollastonite) is ground to a suitable size and is mixed with water to obtain an appropriate water to mineral (W:M) ratio. The slurry is then sent to a pressurized reactor, which is maintained at a particular pressure, and heated to a temperature which is 20 °C less than reactor temperature. Meanwhile, pressurized $CO_2$ obtained from flue gases is let into the reactor chamber. The mineral carbonation process is summarized in Fig. 4.

The carbonation process is influenced chiefly by parameters such as pressure of $CO_2$ (p), temperature and size of mineral particle. The operating values of reactor pressure, water to mineral ratio (W:M), temperature and size of crushed mineral are chosen from an earlier study [11] wherein these process conditions are proven to guarantee maximum energetic efficiency for $CO_2$ sequestration process. The values for the optimum set of conditions are 20 bar ($p_{CO2}$), temperature 200 °C (T), size $d < 38\ \mu m$, and W:M ratio 5 kg/kg which provide 75% carbon capture efficiency to the sequestration process.

The case study assumes sequestration to be performed in the power plant itself as against being located in a far off location. Hence, cost of transportation of $CO_2$ has not been considered.

The $CO_2$ generated is then processed through a capture plant to convert the Calcium mineral, Wollastonite to a carbonate. This process is described by the equation below:

$$C + O_2 \rightarrow CO_2 \quad (1)$$

$$CaSiO_3 + CO_2 \rightarrow CaCO_3 + SiO_2 \quad (2)$$

The Gibbs free energy and enthalpy change for the reaction are $-44$ kJ/mol and $-87$ kJ/mol. Coal contains elements other than carbon such as Sulfur, Nitrogen etc. These elements are oxidized to form the corresponding oxides and along with carbon dioxide form flue gases. No energy penalty is attributed to separation of carbon dioxide from the flue gases.

*A. Assumptions*

The performed calculations are based on assumptions in [11], the carbonation process can be affected by several parameters such as pressure of $CO_2$, pressure of water vapor, size of the feedstock particles, liquid to solid ratio and temperature of reaction. All these parameters play a very crucial role in determining the energy required. The important steps involved in the carbonation process are mentioned below:
  a) Feedstock grinding and crushing
  b) $CO_2$ compression and pumping
  c) Slurry L/S ratio
  d) Carbonate formation

IV. ENERGY ANALYSIS

The energy required for aqueous mineralization of carbon-dioxide can be divided into two categories: Electrical and Heat. The electrical energy is used to perform tasks such as crushing, pumping, compressing etc. The heat energy is used to increase the temperature of the slurry to achieve optimum reaction temperatures [11]. As mentioned previously these energies are very sensitive to several parameters. A particular configuration and the energy requirements for that particular set of conditions are evaluated.

The exothermic nature of the carbonation reaction mentioned in (2) can be used to reduce the energy requirements of the carbonation process. The heat energy can be partially converted into electrical energy through steam engine; this has to do with Carnot Efficiency. Carnot's theorem states that no engine operating between two heat reservoirs can be more efficient than a Carnot engine. The Carnot efficiency is stated below:

$$\text{Carnot Efficiency} = 1 - \frac{T_c}{T_h} \quad (3)$$

Factors such as droplet formation reduce this to numbers much below the Carnot Efficiency. A preliminary analysis of literature confirms that this number is around 50% [34]. This number has been used in the calculations for estimating energy recovered.

As mentioned previously the energy requirements can be calculated as a sum total of electrical and heat energy. For the conditions mentioned in Section III the electrical energy required to capture the amount of $CO_2$ emitted is calculated. The assumptions made in previous section are mentioned here for the sake of completion. Coal plant efficiency is assumed to be 44% and that of a natural gas plant is 55%. The calculations are performed for a 1000 MW fossil fuel powered plant (coal and natural gas). In case of pulverized coal power plant, the total weight of coal used per day is approximately 9815 metric tons. Assuming the percentage of carbon by weight is 60%, the total carbon combusted is 5889 metric tons. Using molar analysis it can be concluded that 21600 metric tons of $CO_2$ is produced each day.

The emission factor ($\alpha$) is defined as the amount of $CO_2$ emitted per kWh of energy produced. In case of pulverized coal power plant this turns out to be 0.9 kg of $CO_2$/kWh.

A similar analysis for a natural gas powered power plant leads to the following numbers: volume of natural gas used daily: $4430 \times 10^3$ m$^3$/day, weight of $CO_2$ produced 7857 metric tons per day, and emission factor of the plant is 0.33 kg of $CO_2$/kWh.

Using numbers mentioned in [11] and above the amount of energy required to reduce the $CO_2$ emissions is computed. The energy for grinding of feedstock is estimated to be around 253 kWh/ton of $CO_2$ captured while the energy required for compression and pumping of $CO_2$ is around 150 kWh/ton of $CO_2$ [11]. The total electrical energy required is equal to 403 kWh/ton of $CO_2$. Heat energy is required to bring the slurry to the optimum reaction temperature (200 °C). The energy required for this stage is 799 kWh/ton of $CO_2$. In addition to these, we can also extract heat from the exothermic reaction of carbon dioxide with calcium silicate. This reaction supplies 752 kWh/ton of $CO_2$ captured. Using these numbers and previous calculations the total energy required is calculated to be $25.96 \times 10^6$ kWh/day. This amounts to a 1.08 GW power plant. A part of this energy can be obtained by the conversion of reaction heat energy into electricity using steam turbine. The amount of electrical energy generated is limited by the steam turbine efficiency. Under such conditions, the electrical energy recovered is $8.12 \times 10^6$ kWh. In other words this is equivalent to a 340 MW plant.

A similar analysis for a natural gas plant mentioned previously results in numbers that are substantially smaller due to the reduced emissions of a natural gas plant. Hence the expectation is reduced energy requirements. The gross total energy required per day is 400 MWh whereas the total energy recovered is 124 MWh, leading to net energy requirement of 276h MW.

The numbers presented in the above section are dependent on many factors, one of them being the emission factor. This directly controls that amount of energy required and recovered. The emission factor ($\alpha$) can go as low as 0.3 kg/kWh for natural gas plants and can be as high as 0.9 kg/kWh for certain coal powered plants. A sensitivity analysis performed using $\alpha$ as the variable and a graph is plotted for the variation of total power required, Fig. 5.

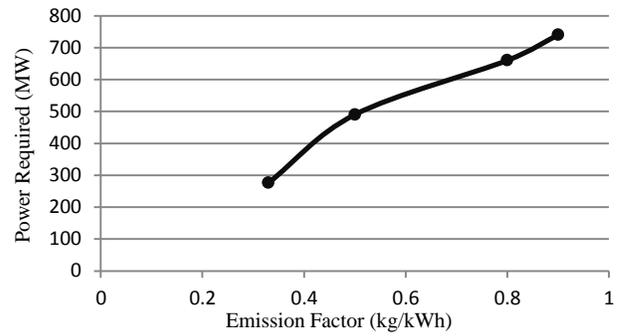

Fig. 5. Total energy required for varying emission factor

Table 1. Summary of power Requirements for carbon capture for natural gas and coal power plants

| Parameters | Pulverized coal plant | Natural gas plant |
|---|---|---|
| Net Fuel consumption per day | 9815 metric tons | $4430 \times 10^6$ lit |
| Emissions of $CO_2$ per day | 21600 metric tons | 7900 metric tons |
| Gross power needed for sequestration | 1080 MW | 400 MW |
| Total power recovered from the reactor | 340 MW | 124 MW |
| Net sequestration power requirement | 740 MW | 276 MW |
| Sequestration power requirement | 74% | 27.60% |

## V. CONCLUSION AND FUTURE WORK

Table 1 indicates that power requirements for carbon capture are a large portion of energy of the power plant for which the capture plant is designed. Results for pulverized coal power plant indicate a power plant that is 42.5% larger is required to capture the $CO_2$ while providing 1000 MW to the end users. The same numbers for natural gas present a slightly less pessimistic outlook. A natural gas power plant needs to be 21.6% larger to supply 1000 MW of electrical output.

As mentioned in equation (2) carbonation reaction is a naturally occurring spontaneous, exothermic reaction. Yet the sequestration process in energy intensive. This is largely due to the slow nature of the reaction. To speed up the kinetics and make the reaction feasible on an industrial scale presents a very good opportunity for research.

The heat recovered from the reactor is used to supply a fraction of the gross power required for sequestration. The energy efficiency of heat to electricity conversion has been considered to be 50% which is the typical case for steam turbines. This intermediate stage presents an attractive opportunity to optimize carbon sequestration energy efficiency. If the reuse of this heat is made more efficient, the net sequestration energy requirement will also reduce.

The purpose of Carbon capture method discussed in this paper is to reduce the amount of $CO_2$ emitted into the atmosphere by power plants. This places tight restrictions on the energy requirements of the process. In principle the mineral carbonation stage needs to be made more energy efficient. Only if such a solution is combined with a highly efficient steam turbine, the implementation discussed here can be treated as a viable solution. Under such circumstances renewable energy sources such as solar, wind, biomass can be considered for meeting the energy needs.